\documentstyle[12pt]{article}
\def\al{\alpha}

\def\ga{\gamma}
\def\de{\delta}

\def\et{\eta}

\def\ka{\kappa}
\def\la{\lambda}

\def\rh{\rho}

\def\si{\sigma}

\def\ta{\tau}

\def\ph{\phi}

\def\Th{\Theta}

\def\mn{{\mu\nu}}

\def\fr#1#2{{{#1} \over {#2}}}
\def\prt{\partial}
\def\ap{\al^\prime}
\def\apt{\al^{\prime 2}}
\def\apth{\al^{\prime 3}}

\def\vev#1{\langle {#1}\rangle}
\def\vev{\langle {\phi}\rangle}

\def\half{{\textstyle{1\over 2}}}
\def\frac#1#2{{\textstyle{{#1}\over {#2}}}}

\def\lsim{\mathrel{\rlap{\lower4pt\hbox{\hskip1pt$\sim$}}
    \raise1pt\hbox{$<$}}}
\def\gsim{\mathrel{\rlap{\lower4pt\hbox{\hskip1pt$\sim$}}
    \raise1pt\hbox{$>$}}}
\def\sqr#1#2{{\vcenter{\vbox{\hrule height.#2pt
         \hbox{\vrule width.#2pt height#1pt \kern#1pt
         \vrule width.#2pt}
         \hrule height.#2pt}}}}
\def\square{\mathchoice\sqr66\sqr66\sqr{2.1}3\sqr{1.5}3}

\newcommand{\beq}{\begin{equation}}
\newcommand{\eeq}{\end{equation}}
\newcommand{\bea}{\begin{eqnarray}}
\newcommand{\eea}{\end{eqnarray}}
\newcommand{\rf}[1]{(\ref{#1})}

\renewenvironment{thebibliography}[1]
 { \rm
   \begin{list}{\arabic{enumi}.}
    {\usecounter{enumi} \setlength{\parsep}{0pt}
     \setlength{\itemsep}{3pt} \settowidth{\labelwidth}{#1.}
     \sloppy
    }}{\end{list}}

\begin{document}
\titlepage

\begin{flushright}
{DAMTP-R92-40\\}
{IUHET 242\\}
{hep-th/9302120\\}
{December 1992\\}
\end{flushright}
\vglue 1cm

\begin{center}
{{\bf CONDENSATES AND SINGULARITIES IN STRING THEORY
\\}
\vglue 1.0cm
{V. Alan Kosteleck\'y$^a$ and Malcolm J. Perry$^b$\\}
\bigskip
{\it $^a$Physics Department\\}
\medskip
{\it Indiana University\\}
\medskip
{\it Bloomington, IN 47405, U.S.A.\\}
\vglue 0.3cm
\bigskip
{\it $^b$D.A.M.T.P.\\}
\medskip
{\it University of Cambridge\\}
\medskip
{\it Silver Street\\}
\medskip
{\it Cambridge CB3 9EW, England\\}

\vglue 0.8cm
}
\vglue 0.3cm

\end{center}

{\rightskip=3pc\leftskip=3pc\noindent
We derive a class of solutions to the string sigma-model equations
for the closed bosonic string.
The tachyon field is taken to form a constant condensate
and the beta-function equations at one-loop level
are solved for the evolution of the metric and the dilaton.
The solutions represent critical string theories in arbitrary dimensions.
The spectrum of the subclass of models with
a linearly rising asymptotic dilaton is found using the Feigin-Fuks method.
Certain approximate solutions arising in string field theory
are used to illustrate the results explicitly.
An argument based on conformal invariance leads to the
conjecture that that stringy corrections
to at least some singular spacetimes in general relativity
result in non-singular metrics.
We use the singularities of the big-bang/crunch type appearing in our models
to examine this conjecture.

}

\vfill
\newpage

\baselineskip=20pt
{\bf\noindent 1. Introduction}
\vglue 0.4cm

Many open issues remain in string theory.
One is the development of solutions
to critical string theories that have non-trivial backgrounds
and another is the description of models in non-critical dimensions.
Part of the motivation for addressing these issues
is the stringy resolution of puzzles in classical
and quantum general relativity,
such as the role and nature of spacetime singularities.
In this paper,
we seek solutions to critical string theories
that describe models in arbitrary dimensions.

Several methods can be used to find string theories
in non-trivial backgrounds.
An important one is the sigma-model approach
\cite{CFMP,FT,L,CKP}.
The idea is to impose conformal invariance on
an effective theory for selected particle fields contained in the string,
by requiring the vanishing of the corresponding beta functions.
The resulting equations can be viewed as arising from
a spacetime action written in terms of the particle fields.
In explicit calculations,
the beta functions are usually determined only at tree-level
in the string perturbation series and at no more than a few orders
in an expansion in powers of the string scale.
Nonetheless, solutions of the resulting expressions are presumably
perturbative approximations to exact solutions of the full string theory
and so may thereby provide insight.

The lowest-lying fields of the closed string are the tachyon,
the dilaton, the graviton, and the Kalb-Ramond field.
Some authors have addressed the issue of the presence of the tachyon
in the beta-function equations.
The proof that the weak-field expansion provides a link between
a suitable tachyon term in the sigma model and string tachyon scattering
amplitudes is given in
\cite{BNY}, while
the beta functions incorporating the tachyon are discussed in
\cite{CG,DS,T}.
However, much of the work on solutions to the beta-function equations
has concentrated on the role of the dilaton and graviton
\cite{MY,ABEN,APS,P,M}.

In generating solutions to the beta function equations
involving some non-zero background values,
it is important to realize that it is mathematically consistent
to neglect the tachyon $T$ only if the tachyon potential $V(T)$
satisfies $V(T)=V^\prime (T)=0$.
Even if these conditions are satisfied,
any resulting solutions are destabilized if $V^{\prime\prime} <0$.
In this paper,
we seek to construct stable string models in background fields
corresponding to arbitrary dimensions.
We consider in particular the case of a constant tachyon expectatation,
$T=T_0$.
This means that
we are seeking simple closed-string analogs of the usual de Sitter,
Minkowski, and anti-de Sitter solutions
arising in a particle field theory when a scalar
forms a Higgs condensate.
These solutions are dynamical,
i.e.,
they represent evolving universes,
in which both the metric and the dilaton fields change with time.

When the action is written in a form as close to the Einstein action
as possible,
the solutions we find have singularities
of the big bang and/or big crunch type.
They provide a means of exploring the second open issue
mentioned above,
namely, whether and how strings permit answers to the
issue of the physical meaning of the
global structure of spacetime in the presence of spacetime singularities.
Here,
we argue that string theory cures
at least some of the spacetime singularities that arise in
general relativity.
The argument is based on the possibility of
performing conformal transformations that smooth away
the singularities.

Provided the asymptotic solution has a linear dilaton background,
information about the spectrum in the sigma model
can be found by an application of the Feigin-Fuks method
\cite{FF}.
Note that a large class of exact solutions
at the string tree level can be found
in terms of WZW models
\cite{WZW},
and various models corresponding to non-zero backgrounds have been constructed
\cite{BS,NW}.
However, our models are not known to fit in this class
\cite{GQ} and so these methods are not of direct use here.

In principle,
an alternative approach to string theories in non-trivial backgrounds
is via string field theory.
For example,
one can seek to construct
an effective potential for chosen particle fields
and then look for non-trivial extrema
\cite{KS2}.
For the field theory of the closed bosonic string in its critical dimension
\cite{KKS,SZ,K},
an analysis of this sort
\cite{KS1}
suggests the existence of stringy solutions of the type we present.
This approximate method of solution of the string-field equations
of motion indicate the existence of an asymptotic
and perturbatively stable vacuum with a non-zero tachyon expectation value.
We use these approximate solutions to illustrate explicitly our
results obtained from the sigma-model approach and the Feigin-Fuks method.

Here is an outline of the contents of this paper.
In section 2,
we set up some formalism and present
to leading level the beta functions
for the tachyon and all the massless fields in the closed bosonic string.
We then describe the basis for the conjecture that
singular spacetimes in general relativity may be non-singular
in the string extension.
In section 3,
we derive a class of solutions of the beta function equations,
valid for any dimension $d$.
They describe cosmologically evolving universes
in which nonzero expectation values appear for
all the particles in the theory except the Kalb-Ramond field.
In section 4,
we use the Feigin-Fuks technique to get information about
the spectrum of the string solutions at asymptotic times
for a subclass of these models.
Approximate results derived from closed-string field theory
are presented in section 5 and used to provide explicit examples.
We summarize and discuss the content of the paper in section 6.

\vglue 0.6cm
{\bf\noindent 2. String $\beta$-functions}
\vglue 0.4cm

The spectrum of the closed bosonic string
in its critical dimension
propagating in flat spacetime with metric $\eta _{\mu\nu}$
is well understood.
At the lowest level, there is a tachyon $T$
with mass $m$, $m^2=-4/\alpha'$.
The first-excited states are massless and consist of
a graviton $h_{\mu\nu} = h_{\nu\mu}$,
a dilaton $\ph$	and a Kalb-Ramond field $B_{\mu\nu}= - B_{\nu\mu}$.
Above these are an infinite number of massive fields.
For non-critical string theory in spacetime dimension $d$,
the tachyon mass is given by $m^2=-(d-2)/6\alpha '$.
The first-excited state remains massless but the graviton, dilaton and
Kalb-Ramond fields appear as discrete states.
Note that $d\le 26$ seems to be demanded by the no-ghost theorem;
see, for example,
ref. \cite{CT}.

Each of the fields
$T$, $h_{\mu\nu}$, $B_{\mu\nu}$ and $\phi$ are long range and
thus may reasonably be expected to form a condensate in spacetime.
The other fields all have large masses,
and so their direct influence on physics
at large scales should be minimal.
We can construct the action $I$
for a string propagating in a background where
the long-range fields do not vanish.
Let the string world sheet $\Sigma$ have a metric $\gamma _{ab}$
and coordinates $\xi^a$, $a=1,2$.
Let the location of $\Sigma$ in spacetime
be given by $X^\mu (\xi ^a)$.
The action $I$ is then
\beq
I=- {1\over 2\pi\alpha'} \int _\Sigma d^2\xi\ \sqrt{\gamma}\left( \half
    \gamma ^{ab}\partial_a X^\mu \partial_b X^\nu g_{\mu\nu} +
   \half \epsilon ^{ab}\partial _a X^\mu \partial_b X^\nu B_{\mu\nu}
   +\half \ap T -\frac 1 4 \ap R^{(2)}\phi\right) .
\label{ba}
\eeq
Here, $g_{\mu\nu}$ is the background metric of a curved spacetime,
coming from the combination of $h_{\mu\nu}$ and $\eta_{\mu\nu}$.
The symbol $\epsilon ^{ab}$ represents the alternating tensor on $\Sigma$
and $R^{(2)}$ is the Ricci curvature scalar
formed from the metric $\gamma_{ab}$.
Note that $\xi^a = (\ta, \si)$, $\ga_{ab}$, $g_{\mu\nu}$, $B_{\mu\nu}$,
the dilaton $\ph$, and the tachyon $T$ are dimensionless,
while $X^\mu$ has length dimension one.
Various terms involving fields for higher-mass excitations
of the string could be added to $I$,
but they would lead to apparently unrenormalizable interactions.

The physical significance of $\phi$ can be understood from the action $I$.
If $\phi$ is constant over $\Sigma$,
the contribution to the action from the dilaton term becomes
\beq
I_{\phi}=\langle\phi\rangle {1\over 8\pi}\int_\Sigma d^2\xi
\ \gamma^{1/2} R^{(2)}
=\half \langle\phi\rangle \chi
\quad ,
\label{bc}
\eeq
since the integral over the string world sheet
is just the Gauss-Bonnet invariant,
which is related to the Euler character $\chi$ of the world sheet.
A world sheet of genus $g$ with $h$ ends has an Euler character of
$\chi=2-2g-h$.
Thus,
the effect of adding an extra end to the world sheet
is to give an additional contribution to
the path integral of $exp(\half\langle\phi\rangle)$.
This means that the string coupling constant is
\beq
\hat g=exp(\half \vev )
\quad .
\label{bd}
\eeq
Note that nothing unphysical happens if
$\langle\phi\rangle \to -\infty$.
The string coupling constant
merely goes to zero, so free string theory yields the only physical
contributions to the path integral.
Conversely, if $\langle\phi\rangle \rightarrow \infty$
then strong string coupling occurs,
and our picture of physics based on string perturbation
theory breaks down.

The basic symmetry of string physics is world-sheet conformal invariance.
Classically,
this symmetry is violated in $I$
by the tachyon and dilaton fields.
Nevertheless, it can be restored quantum mechanically.
As is apparent from the form of $I$,
the background spacetime fields $g_{\mu\nu}, B_{\mu\nu}$, and $\phi$
behave like coupling constants
and the field $T$ behaves like a world-sheet cosmological constant.
The quantum-mechanical statement of conformal invariance is that the
$\beta$ functions for all these fields vanish.
Evaluating the $\beta$ functions to one loop gives
\beq
R_{\mu\nu} = \nabla_\mu\nabla_\nu \phi + \nabla_\mu T\ \nabla_\nu T +
\frac 1 4 H_{\mu\rh\si} H_\nu ^{\phantom{\nu}\rh\si}
\quad
\label{be}
\eeq
for the metric,
\beq
\nabla_\la H^{\la\mu\nu} + (\nabla_\la \phi) H^{\la\mu\nu} =0
\quad
\label{bf}
\eeq
for the Kalb-Ramond field,
and
\beq
\square\ T + (\nabla_\mu\phi ) (\nabla^\mu T) = V'(T)
\quad
\label{bg}
\eeq
for the tachyon.
In these equations,
$H_{\la\mu\nu}= 3\partial_{[\la}B_{\mu\nu]}$
is the 3-form field strength found
from the exterior derivative of $B_{\mu\nu}$.
Also,
$V'(T)= \prt V/\prt T$ is the derivative of the
tachyon potential $V(T)$,
the explicit form of which is discussed in section 5.
The beta function for the dilaton can also be calculated.
Requiring its vanishing to the same order as Eqs. \rf{be} - \rf{bg} gives
\beq
\hat c = R-(\nabla\phi)^2
- 2\square\ \phi -(\nabla T)^2 - 2V(T)
-\frac 1 {12} H_{\la\mu\nu} H^{\la\mu\nu}
\quad ,
\label{bi}
\eeq
where
$\hat c = 2(d-26)/3\alpha'$.
In the absence of background fields,
$\hat c$ would be the central charge $c$ divided by $3 \ap /2$;
see ref. \cite{NY}.

All these conditions can be derived from the spacetime action
\beq
I_{S}= - \fr 1 {2\ka^2} \int d^d x \sqrt g e^\phi \left( \hat c - R -
(\nabla\phi)^2 + (\nabla T)^2 + 2V(T) + \frac 1 {12}
H_{\la\mu\nu} H^{\la\mu\nu}\right)
\quad ,
\label{bj}
\eeq
where $2\ka^2 = 16\pi G_N$ when $d=4$.
Thus, we can derive the stringy version of the Einstein
equation and the relevant wave equations from a symmetry principle.

Conformal invariance on the string world sheet
induces conformal covariance in spacetime.
One can always redefine the metric
by conformal transformations involving the dilaton.
Thus, under conformal transformations the mapping
\beq
g_{\mu\nu}  \mapsto  f(\phi)g_{\mu\nu}~~,~~~~
B_{\mu\nu} \mapsto B_{\mu\nu}~~,~~~~
T \mapsto T~~,~~~~
\phi \mapsto \phi
\quad
\label{bk}
\eeq
is an invariance of the system.
Nonetheless,
a conformal transformation of the form \rf{bk}
induces non-trivial changes in
both the $\beta$ functions and the action
because both the connection coefficients
and the curvature transform in a non-trivial way.

The action \rf{bj} is expressed in the `string' frame,
so named because
this form arises naturally in the $\sigma$-model approach to string theory.
However, to discuss spacetime physics it is useful to write the
action in a form as close to the Einstein action as possible.
To arrange this,
choose
\beq
f(\phi)= e^{-2\phi/(d-2)}
\quad ,
\label{bl}
\eeq
which has the effect of removing the factor $e^\ph$ multiplying $R$
in Eq. \rf{bj}.
The action in the `Einstein' frame is then
\bea
I_{\rm E} = - \fr 1 {2\ka^2} \int _M d^d x \sqrt g
\bigl[ -R & + & \hat c e^{-2\phi/(d-2)}
+{1\over d-2}(\nabla\phi)^2 + (\nabla T)^2 \nonumber \\
& + & 2 e^{-2\phi/(d-2)}V(T)
+ {1\over 12} e ^{4\phi/(d-2)}H_{\la\mu\nu} H^{\la\mu\nu}\bigr]
\quad .
\label{bm}
\eea
Note that the graviton is not directly coupled to the dilaton in this frame.

The question arises as to whether one frame is more `natural' than another.
Evidently,
if one wishes to compare string predictions with general relativity,
it is useful to use the Einstein frame.
In contrast, from the sigma-model viewpoint the `string' frame appears
more natural.
We believe neither frame should be taken as fundamental in the real sense
of the word.
Conformal invariance ensures
that the choice of frame is a matter of convenience
in the description.

It is nonetheless true
that the arbitrariness of $f(\phi)$ in this formulation
affects our view of the physics near singularities.
Consider first the situation in general relativity,
where spacetime singularities are viewed as the boundary of spacetime.
They are usually related to singularities of the metric tensor.
However, the converse is not true:
a metric singularity may be a coordinate singularity
or it may be a spacetime singularity.
The way to discern the difference is to seek a
diffeomorphism to a new, nonsingular metric.
If one exists, the singularity was a coordinate artifact.
We view the situation concerning conformal transformations
in string theory to be analogous.
Locally,
rewriting the $\beta$ function in a new
conformal frame merely maps the solutions of the equations
into new solutions.
Globally, the situation is radically different.
It might be the case that a particular spacetime appears singular
in some coordinate frame.
If we can find a $f(\phi)$ removing the singularity,
then a non-singular conformal frame exists.
We regard combinations of the dilaton field
and the metric where this can be done as non-singular.
This situation therefore allows us
to view as non-singular many spacetimes
that would be singular in general relativity.
In this way, string theory may resolve one of the
perplexing issues of general relativity.
This might seem strange from the point of view of spacetime physics,
where at a singularity it is impossible to solve differential equations.
However,
the spacetime is determined by the string behavior,
which may be well-defined
even though the spacetime picture suffers from pathologies.

To remove a spacetime singularity via such a transformation,
$f(\phi)$ itself must be singular.
If this happens because $\phi$ is tending to $-\infty$,
then nothing odd happens to string physics
because this merely indicates that the string coupling is
approaching zero, cf. Eq. \rf{bd}.
However,
if $\phi$ is tending to $+\infty$
then perturbation theory will break down,
and a more sophisticated treatment of the string may be needed.

\vglue 0.6cm
{\bf \noindent 3. Solutions of the $\beta$-function Equations}
\vglue 0.4cm

In this section,
we address the issue of some solutions
to the $\beta$-function equations
other than Minkowski spacetime in the critical dimension.
One well-known solution for $d>26$ is the linear dilaton
background in which spacetime is flat,
$T=0$, and the dilaton rises linearly with time \cite{P}.
Examination of the dilaton beta function \rf{bi} shows that
\beq
\ph (t) = \mu t~~,~~~~\mu = \sqrt{\fr {2(d-26)}{3 \ap}}
\quad .
\label{ca}
\eeq
This solution is still plagued by a tachyon,
as can be seen by the following argument.
Consider the linear part of the tachyon beta function:
\beq
\square\ T + (\nabla_\la\phi ) (\nabla ^\la T) = - \fr 4 \ap T
\quad .
\label{cb}
\eeq
The tachyon is minimally coupled to the dilaton.
The tachyon mass can be determined from this equation by eliminating
the term in $\nabla T$ using \rf{ca}.
Thus, Eq. \rf{cb} becomes
\beq
\bigl [ - (\prt_t + \half \mu )^2 + \nabla^2 \bigr ] T
= - \left( \fr {4}{\ap}	+ \fr {\mu^2}{4} \right) T
\quad
\label{cc}
\eeq
and the physical mass of the tachyon field is
\beq
m^2 = \fr {-d+2}{6\ap}
\quad .
\label{cd}
\eeq
This calculation agrees with the standard calculation based
either on the Casimir energy of the ground state
or on the anomaly in the Virasoro algebra.

The idea we explore here is to suppose that the tachyon potential
has extrema other than the one at $T=0$.
At $T=0$,
$V(T)$ is a maximum and so the excitations of the field
$T$ about the origin are interpreted as tachyons.
In general,
the mass $m$ of the field that represents small fluctuations about
some fixed value $T_0$ of $T$ is given by $m^2=V''(T_0)$,
where $V'(T_0)=0$.
Thus,
for $T_0=0$ we find $m^2=-4/\alpha '$.
In this section,
we assume that $V(T)$ has an extremum at $T=T_0 \neq 0$,
with $V''(T_0)\geq 0$ to avoid problems with stability and causality.

In field theory,
a tachyon can appear as a consequence of
selecting an unstable vacuum state,
but disappears after spontaneous symmetry breaking has occurred.
In this case,
a non-zero value of $V(T_0)$
gives rise to an effective cosmological constant.
The vacuum state of the resulting theory is
de Sitter space if $V(T_0)>0$,
Minkowski space if $V(T_0)=0$,
or anti-de Sitter space if $V(T_0)<0$.
In contrast, in string theory
the vacuum state cannot be so simple because $V(T)$ is coupled
to the gravitational field through the dilaton.
As is apparent from the dilaton beta function,
any curvature causes the dilaton to vary over spacetime.
This means the effective cosmological constant
can depend on the location in spacetime.
We therefore expect to find string solutions
that are more complicated than de Sitter,
Minkowski, or anti-de Sitter spaces.

A plausible ansatz is to suppose
that spatial sections of the spacetime metric
are flat $R^{d-1}$ coordinatized by $(x_1,x_2,\dots ,x_{d-1})$.
The assumption of flatness is reasonable
because the false-vacuum configuration corresponding to
$T_0=0$ can evolve into a new configuration without any topology change.
We therefore seek a spacetime described by a $k=0$
Friedman-Robertson-Walker-Lema\^{\i}tre universe with line element
\beq
ds^2 = -dt^2 + a^2(t)(dx_1^2 + dx_2^2 + \dots dx_{d-1}^2)
\quad .
\label{ce}
\eeq
Here, $t$ is called the cosmic time coordinate
because it is the same as proper time
for geodesic observers at constant $x_i$.
If $a(t)=e^{Ht}$ the spacetime is de Sitter,
if $a(t)$ is constant the spacetime is Minkowski,
and if
$a(t)=\cos Ht$ the spacetime is anti-de Sitter.
Whatever the functional form of $a(t)$,
these spacetimes are conformally flat.
This can be seen by defining a conformal time co-ordinate $\eta$ by
\beq
\eta = \int {dt\over a(t)}
\quad ,
\label{cf}
\eeq
so that
\beq
ds^2 = a^2(\eta) (-d\eta ^2 + dx_1^2 + dx_2^2 + \dots dx _{d-1}^2)
\quad .
\label{cg}
\eeq
In what follows,
for simplicity
we also assume that the dilaton $\phi$ and the tachyon fields are
functions only of $t$,
and we take $H_{abc} = 0$.

The $\beta$-function equations can now be written out explicitly.
In the string frame,
the 00 and $jj$ components of the beta function \rf{be}
for the metric are
\beq
(d-1){\ddot a\over a} + \ddot \phi + \dot T^2 =0
\quad
\label{ch}
\eeq
and
\beq
a\ddot a + (d-2) \dot a^2 + a\dot a\dot \phi = 0
\quad ,
\label{ci}
\eeq
respectively.
The $0j$ components identically vanish.
Equation \rf{bg} for the tachyon becomes
\beq
\ddot T + (d-1) {\dot a\over a} \dot T + \dot \phi \dot T + V'(T)
=0
\quad .
\label{cj}
\eeq
Finally,
Eq. \rf{bi} for the dilaton is
\beq
\dot \phi ^2 + \ddot \phi + (d-1){\dot a\over a}\dot \phi -2V(T) =\hat c
\quad .
\label{ck}
\eeq
In all these equations,
a dot over a field represents the $t$ derivative
and a prime represents the $T$ derivative.

We can integrate these equations directly.
If $V'(T)=0$,
then $T=T_0$ is clearly a solution to Eq. \rf{cj}.
If we now put $h=\ln (\sqrt{\ap}~\dot a/a)$ and eliminate $\phi$
from Eqs. \rf{ch} and \rf{ci},
we find
\beq
\ap \ddot h=(d-1) e^{2h}
\quad .
\label{cl}
\eeq
The first integral of this equation is
\beq
\dot h^2 = \fr {(d-1)} \ap e^{2h} + k
\quad .
\label{cm}
\eeq
Here, $k$ is an integration constant with dimension $1/\ap$.
Solutions to this equation lie in three distinct classes
depending on whether the integration constant $k$
is positive, negative or zero.
For each case,
we can then determine $\phi$ and examine Eq. \rf{ck}
to obtain the value of $c$ to which the solution corresponds.
Note that applying the Bianchi identities to \rf{ch} - \rf{cj}
implies that \rf{ck} is an identity modulo the determination of $\hat c$.

Suppose initially $k=0$.
Two solutions for $a(t)$ arise:
\beq
a(t) = a_0t^{1/\lambda}~~,~~~~
\phi (t) = \phi _0 + (1-\lambda)\ \ln (t/\sqrt{\ap})
\quad
\label{cna}
\eeq
and
\beq
a(t) = a_0t^{-1/\lambda}~~,~~~~
\phi (t) = \phi _0 + (1+\lambda)\ \ln (t/\sqrt{\ap})
\quad .
\label{cnb}
\eeq
Here,
$\lambda =(d-1)^{1/2}$, and $a_0$ and $\phi _0$ are arbitrary
constants of integration.

Similarly, for $k>0$ we find two solutions:
\bea
a(t) & = & a_0 (\tanh \half\la t\sqrt{k})^{1/\la}
\nonumber \\
\ph (t) & = & \phi _0 + (1 + \la)\ln \,\cosh \half\la
\sqrt{k}\, t + (1 - \la) \ln \,\sinh \half\la \sqrt{k}\, t
\quad
\label{coa}
\eea
and
\bea
a(t) & = & a_0 ( \tanh \half\la t\sqrt{k})^{-1/\la}
\nonumber \\
\phi (t) & = & \phi _0 +
(1-\la)\ln\,\cosh\half\la \sqrt{k}\ t
+ (1+\la)\ln\,\sinh\half\la \sqrt{k}\ t
\quad .
\label{cob}
\eea
For small $t$,
these solutions are the same as those with $k=0$.
As $t$ becomes large compared to $1/\lambda \sqrt{k}$,
we see that $a(t)\to a_0$ and $\phi (t)\to \lambda \sqrt{k}\ t$.
Thus,
for large $t$ these solutions tend asymptotically to flat space
with a dilaton field that is rising as a linear function of $t$.

For $k<0$,
again there are two solutions:
\bea
a(t) & = & a_0  (\tan \half\la\sqrt{-k}\, t)^{1/\la}
\nonumber \\
\phi (t) & = & \phi _0 + (1 - \la) \ln\,\sin\half\la\sqrt{-k}\, t
+ (1+\la)\ln\,\cos\half\la\sqrt{-k}\,t
\quad
\label{cpa}
\eea
and
\bea
a(t) & = & a_0  (\tan \half\la\sqrt{-k}\, t)^{-1/\la}
\nonumber \\
\phi (t) & = & \phi _0 + (1+\la)\ln\,\sin\half\la\sqrt{-k}\, t
+ (1-\la)\ln\,\cos\half\la\sqrt{-k}\,t
\quad .
\label{cpb}
\eea
Again, for small $t$ these have the same behavior as the $k=0$ case.
However,
we see that $a(t)$ diverges
as $t\to\pi/\la\sqrt{-k}$ in \rf{cpa}.
Thus, in this solution
the universe tends to infinite size in a finite amount of proper time,
which presumably is unphysical.
In contrast,
in \rf{cpb} the universe starts out
with infinite size and then collapses,
with the big crunch at $t=\pi/\lambda\sqrt{-k}$.
This case appears dynamically unstable.

Substitution of all these solutions into \rf{ck} yields
\beq
\hat c+2V(T_0) = (d-1) k
\quad .
\label{cq}
\eeq
For positive, zero, and negative values of $k$,
the solutions are analogs of the de Sitter,
Minkowski, and anti-de Sitter solutions in field theory, respectively.

In all these solutions,
the spacetime is singular whenever $a(t)\to 0$.
The universe then has zero size,
corresponding either to the big bang or to the big crunch.
However, this singularity occurs in the string frame.
While local physics can be translated between one frame and another in a
predictable way,
the presence or absence of singularities involving the global structure
is not straightforward to translate.
Since we can pass from one frame to another
via conformal transformations of the form $e^{\alpha\phi}$,
$\alpha\in \bf R$,
the spacetime may well be non-singular in one frame
but singular in another.
As an example,
consider the solution \rf{cna}.
Here, it is best to transform to conformal time.
{}From \rf{cf} we find
$\eta\sim t^{1-1/\lambda}$
so that
$a(\eta)\sim\eta ^{1\over \lambda -1}$.
Since $\phi  (t) =\phi _0 + (1-\lambda)\ \ln (t/\sqrt{\ap})$,
we find $e ^\phi \sim \eta ^{-\lambda}$
and hence $a(\eta)\sim e ^{-\phi\ {1\over\lambda(\lambda -1)}}$.
Thus,
if we conformally rescale the metric for \rf{cna}
by a factor $e^{2\phi/\lambda(\lambda -1)}$,
the spacetime becomes flat and hence non-singular.

It is natural to ask what kinds of singularities are allowed in general.
One possible criterion,
suggested by analogy with field theory,
is the integrability of the action density at $t=0$.
In the present simple example,
$e^\phi \sim t^{1-\lambda}$
and
$g^{1/2}\sim a(t) ^{d-1}\sim t^{d-1/\lambda}$
so
$g^{1/2} e^\phi\sim t^{1-\lambda + {d-1\over \lambda}}$.
Substituting \rf{ck} into the action in the string frame gives
\beq
I_S= \fr 1 {\ka^2} \int d ^d x\, g^{1/2} e^\phi
\left(\square\phi + (\nabla\phi)^2\right)
\quad .
\label{cr}
\eeq
Using the functional form of $\phi$ and the metric then shows there is
no singular contribution to $I_S$.

This suggests that the singularities found in string metrics are harmless,
unlike similar-seeming ones found in general relativity
representing the boundary of spacetime.
In our example,
there is apparently no obstacle to passing through the singularity
to some other region of spacetime.
It is attractive to conjecture that this picture
holds generically,
in which case the issue of singularities
in general relativity is resolved by string theory.

\vglue 0.6cm
{\bf \noindent 4. Feigin-Fuks Treatment for $k>0$}
\vglue 0.4cm

The consistency of our calculations can be checked
for the specific case where $k>0$.
For large times $t \gg 1/\la\sqrt{k}$,
spacetime is flat and the dilaton is growing linearly
in time as $\ph \sim \la \sqrt{k} t$
in both of the cases \rf{coa} and \rf{cob}.
It is possible to construct the first-quantized string exactly
in such a background by using the Feigin-Fuks method
\cite{FF}.
Starting from the action \rf{ba} in this background
and assuming that the tachyon condensate causes the tachyon
to have a constant expectation value $T_0$,
we discover that the energy-momentum tensor $\Th_{ab}$
of the string is given by
\bea
\Th_{ab} = \prt_a X^\mu \prt_b X^\nu \et_\mn
& - & \half \ga_{ab} \prt_c X^\mu \prt_d X^\nu \et_\mn \ga^{cd}
-\half \ap T_0 \ga_{ab}
\nonumber \\
& + &\half \ap \la \sqrt{k}
(\nabla_a \nabla_b X^\mu - \ga_{ab} \square X^\mu) \de_\mu^0
\quad ,
\label{da}
\eea
where $\et_\mn$ is the target-space metric with signature $(-+++\ldots)$.
Note that this tensor is \it not \rm trace free:
the contributions of $T$ and $\ph$
to the action are not conformally invariant.
The present calculation is semiclassical.
Only at the quantum level (discussed below) is full conformal invariance
restored.

The equations of motion of the string are
\beq
2\ga^{ab}\nabla_a \nabla_b X^\mu
= -\half \ap \la \sqrt{k} R^{(2)} X^0 \de^\mu_0
\quad
\label{db}
\eeq
where $R^{(2)}$ is the Ricci scalar of the world-sheet metric.
{}From this it follows that the quantization of the $X^\mu$
other than $X^0$ is unaffected by the time-dependent dilaton.
Note that the tachyon condensate $T_0$ behaves like a
world-sheet cosmological constant.

Let us next consider a flat-cylindrical world sheet
in the orthonormal gauge,
chosen so that the time coordinate $\ta$ is free
and the space coordinate $\si$ varies from 0 to $\pi$.
Solving the string equations of motion
yields the usual expression
\beq
X^\mu = x^\mu +2 \ap p^\mu \ta +i\sqrt{\half\ap}
\sum_{n\ne 0} \left[ \fr {\al^\mu_n} n exp[-2in(\ta - \si)]
+ \fr {\bar\al^\mu_n} n exp[-2in(\ta + \si)]\right]
\quad .
\label{dc}
\eeq
Here, $x^\mu$ is the string center-of-mass coordinate,
$p^\mu$ is the momentum,
and $\al_n$ and $\bar\al_n$ are the usual creation and annihilation
operators for the right- and left-moving excitations, respectively.
As usual,
quantization imposes the commutation rules
\beq
[x^\mu, p^\nu] = i \et^\mn ~~,~~~~
[\al^\mu_m, \al^\nu_n] = [\bar\al^\mu_m, \bar\al^\nu_n] =
m \de_{m+n,0}\et^\mn
\quad .
\label{dd}
\eeq
The ghost sector can be treated similarly.
In what follows,
we concentrate on the differences between our situation
and the usual zero-condensate case.

The (dimensionless) hamiltonian is proportional to the integral of $\Th_{00}$
over a spatial section of the string and is classically given by
\beq
H = \sum_{n=1}^\infty (\al_n \al_{-n} + \bar \al_n \bar\al_{-n})
+ \ap p^2 + \fr {T_0}{4}
\quad .
\label{de}
\eeq
The Virasoro operators are the Fourier components of the energy-momentum
tensor,
which are modified from their usual form
by the dilaton and tachyon condensates.
For the right movers,
they are
\beq
L_m = \half \sum_{n=-\infty}^\infty \al_n \al_{m-n}
-i\la\sqrt{\fr {\ap k} 8} m \al_m^0
+ \fr {T_0} 4 \de_{m,0}
\quad ,
\label{df}
\eeq
where we have defined
$\al_0^\mu = \sqrt{\half\ap} p_0^\mu$.

At the quantum level, however,
we must deal with the normal-ordering problem.
We therefore replace the $L_m$ by their quantum analogs
\beq
L_m = \half \sum_{n=-\infty}^\infty : \al_n \al_{m-n} :
-i\la\sqrt{\fr {\ap k} 8} m \al_m^0
+ \fr {T_0} 4 \de_{m,0} -A\de_{m,0}
\quad ,
\label{dg}
\eeq
where the colons indicate normal ordering.
The constant $A$ corresponds to the normal-ordering shift
in the vacuum energy,
which can be evaluated by zeta-function regularization.
Its value is then $A = d/24$.
An expression similar to \rf{dg} holds for the left movers.

It can be checked that the Virasoro algebra is still satisfied,
but the central term becomes modified.
Explicitly,
\beq
[L_m, L_n] = (m-n) L_{m+n}
+ (\fr d {12} - \fr {\ap k \la^2} 8 )m^3 \de_{m+n,0}
- \fr d {12} m \de_{m+n,0}
\quad .
\label{dh}
\eeq
The coefficient of $m^3 \de_{m+n,0}$
is the coefficient of $R^{(2)}$ in the trace anomaly,
that is, the dilaton beta function,
while the coefficient of $m\de_{m+n,0}$ is the
coefficient arising from the Casimir energy.
The ghost contributions
can now be incorporated in the usual way.
The condition for vanishing cubic term in the central charge
is therefore
\beq
\fr {d - 26} {12 \ap} = \fr {k \la^2} {8}
\quad .
\label{di}
\eeq
This expression should be the same as the condition
for the vanishing of the dilaton beta function.
Note that Eq. \rf{di} is consistent with Eq. \rf{ca}
if $\mu^2$ is identified with $k\la^2$.

However,
there are quantum corrections to this formula from contributions
of order $(\ap)^0$.
These have been calculated in the dilaton beta function.
In effect,
the Feigin-Fuks construction has evaluated the contribution to the
dilaton beta function by a different route.
The advantage of this method is that it also allows us to make
statements about the string spectrum.
We thus deduce from Eq. \rf{bi} that for consistency to lowest order
\beq
\fr {d - 26}{12 \ap} - \fr {k \la^2} {8} = -\fr {V(T_0)} {4}
\quad .
\label{dj}
\eeq
Note that this equation is the same as Eq. \rf{cq}.
This means that we have found a consistent solution
if $V(T)$ has an extremum with $V(T_0)>0$,
provided $k>0$.

At this stage,
we can address the question of whether there is a tachyon
in the spectrum of the string.
The world-sheet hamiltonian must vanish.
The Casimir energy of the world-sheet particles can be found as follows.
The contribution from the $d$ free bosons is $-d/24$,
while the world-sheet reparametrization ghosts provide
a factor $1/12$.
Thus, from \rf{de} we find
\beq
\ap p^2 + \fr {T_0}{4} = \fr {d-2}{6}
\quad .
\label{dk}
\eeq
The mass of the ground state is therefore given by
\beq
m^2 = \fr {4-2d+3 T_0}{12\ap}
\quad ,
\label{dl}
\eeq
and $m^2 \ge 0 $ if $T_0 \ge 2(2-d)/3$.

\vglue 0.6cm
{\bf \noindent 5. String Field Theory and the Tachyon Potential}
\vglue 0.4cm

The discussion in the preceding sections has been in the context
of first quantization.
Another way of obtaining information about background structure
is to use string field theory.
Given a particle field $f$ in a string field theory,
we can find possible condensates of $f$ by
constructing an effective potential for $f$
and looking for extrema.
The physics is then determined by
small oscillations of the particle fields about a particular extremum.
For example,
in this way we can find, at least in principle,
the spectrum of the string theory in a non-trivial background.
This procedure is analogous to that in ordinary field theories.
For instance,
the electroweak model in the naive first-quantized vacuum has
massless vectors and a tachyon $h$,
called the Higgs field.
Interactions in the field theory generate a potential for $h$,
with minima away from the origin in which a Higgs condensate is favored.
If we look at small oscillations in the background Higgs field,
we see that the physics is changed:
there is no tachyon, and massive vector bosons appear.

The field-theoretic approach
has an advantage over first-quantized methods in that
issues about the meaning or existence of vacuum structure are avoided.
The point is that
string field theory purports to be a consistent off-shell formulation,
and as such provides a definite off-shell framework within which to
examine the question of a non-trivial background.
In practice, however,
the presence in string field theory of an infinite number of particle fields
and, especially for the closed string,
the complexities of the interactions
makes an exact treatment along these lines difficult.
Instead,
approximation methods are needed.

One approximation scheme is presented in
refs. \cite{KS2,KS1},
which examined the issue of the formation of condensates
in string field theories in their critical dimension
using the level-truncation method.
In particular,
ref. \cite{KS1} obtained an approximation to a stable ground state
of the closed bosonic string via the following path.
In the particle-field expansion of the closed-string field
about a flat background in 26 dimensions,
apply the truncation at the first-massive level,
thereby keeping the tachyon, dilaton, graviton, and Kalb-Ramond fields
but disregarding all higher-mass fields.
For simplicity,
restrict attention to the cubic term in the nonpolynomial action
\cite{KKS,SZ,K} and choose the Siegel-Feynman gauge \cite{SI},
which eliminates certain auxiliary fields for convenience.
At this stage,
one is left with about 50 interaction terms in the lagrangian density.
Presumably,
an ansatz of the form \rf{ce} would lead to solutions including
those we presented in section 3.
However,
most of these interactions are derivative couplings,
which can be disregarded if an approximation with only stable ground states is
sought.
The few remaining terms form the static potential.
The usual 26-dimensional vacuum is an unstable extremum of this potential,
in which the background string field is zero.
At this level of approximation another extremum exists,
in which the tachyon acquires a constant nonzero expectation.
Condensates of the dilaton, graviton, and Kalb-Ramond fields do not appear.

The tachyon potential $\hat V(\hat T)$ appearing in the lagrangian
of the string field theory has mass dimension 26 and takes the form
\beq
\hat V(\hat T) = - \fr 2 \ap \hat T^2 + \fr {\bar g}{3!}\hat T^3 + \ldots
\quad .
\label{ea}
\eeq
Here,
$\hat T$ represents the tachyon field (with mass dimension 12)
appearing in the expansion of the closed-string field,
and $\bar g$ is the tree-level
three-tachyon coupling defined at zero momentum.
The latter is related to the on-shell tree-level three-tachyon coupling $g$
by $\bar g = 3^9 g/2^{12}$.
Extremizing the first two terms in $\hat V$ shows that,
in addition to the local maximum at $\hat T = 0$,
there is a minimum at $\hat T = \hat T_0 = 8/\bar g \ap$
of depth $\hat V(\hat T_0) = - 2^7/3 \bar g^2 \apth $
and curvature $\hat V^{\prime\prime}(\hat T_0) = 4/\ap$.
For definiteness in what follows,
we neglect the quartic and higher terms in this expression.
Note, however,
that in principle terms involving arbitrary powers of $\hat T$
are of importance in determining the exact tree-level potential \cite{KS3}.

In previous sections,
we have worked with a dimensionless tachyon $T$
and the tachyon potential $V(T)$ with mass dimension two.
The connection between these and $\hat T$, $\hat V$ is given by
\beq
T = \ka \bar g \ap \hat T ~~, ~~~~ V(T)
= \ka^2 \bar g^2 \apt \hat V(\hat T)~~,~~~~\ka = \fr 1 4
\quad ,
\label{eb}
\eeq
so that
\beq
V = - \fr 2 \ap T^2 + \fr 1 {3! \ka\ap} T^3 + \ldots
\quad ,
\label{ec}
\eeq
with a minimum at $T = T_0 = 8\ka$
of depth $V(T_0) = - 2^7\ka^2/3\ap$
and curvature $V^{\prime\prime}(T_0) = 4/\ap$.

We can use these approximate expressions to obtain explicit results
from the formulae found in sections 3 and 4.
{}From Eq. \rf{cq},
we find
\beq
k = \fr {2(d-26-128 \ka^2)}{3(d-1)\ap}
\quad .
\label{ed}
\eeq
With $\ka = \fr 1 4$,
this would suggest the presence of a nonzero tachyon condensate
causes the critical dimension to shift from 26 to 34.
It is possible that modular invariance could be recovered
for this case since 34 is a special case of $8n+2$ \cite{TM}.
Note also that for $d = 26$ we have $k = -256\ka^2/75\ap <0$.

{}From the Feigin-Fuks analysis of section 4,
we find the lowest-lying state in the theory using Eq. \rf{dl}.
Using the minimum at $T_0 = 8$ gives
\beq
m^2 = \fr {2+12\ka -d}{6\ap}
\quad .
\label{ee}
\eeq
With $\ka = \fr 1 4$,
this suggests the existence of a string theory without
massless states, and hence infrared finite,
for $d \le 4$.

\vglue 0.6cm
{\bf \noindent 6. Discussion}
\vglue 0.4cm

In this paper,
we demonstrate that the closed bosonic string in its critical
dimension has solutions
that describe critical strings in arbitrary dimensions.
The change in dimensions is accomplished by
balancing the tachyon condensate against the
central charge.
We derive solutions that satisfy the leading-order
terms in the beta-function equations
in an expansion in powers of $\ap$.
As usual,
the higher-order corrections to these equations
are expected to induce higher-order corrections
in the solutions,
which are calculable at least in principle.

The solutions we find are presented in Eqs. \rf{cna} to \rf{cpb}.
They fall into three classes characterized by the sign
of a parameter $k$.
This situation is reminiscent of the de Sitter, Minkowski,
and anti-de Sitter solutions that appear in general relativity
when a Higgs condensate forms.
The parameter $k$ controls the balance between the
central charge and the energy density in the tachyon condensate,
cf. Eq. \rf{cq},
and so the value of $k$ determines the dimensionality of the
string theory involved.
For $k>0$,
the dilaton is asymptotically rising,
which means the Feigin-Fuks method can be applied to
determine the spectrum of the theory.
We use this to find
the value of the tachyon potential at the tachyon expectation value
in terms of $k$
(Eq. \rf{dj})
and a condition that determines
whether the ground state in the background configuration of fields
has positive mass
(Eq. \rf{dl}).
The operator and beta-function methods give results in agreement
where they can both be used.

An approximate solution can be found
to the string field theory of the closed bosonic string
that describes a stable background with constant tachyon expectation value.
In section 5,
we use this approximation to illustrate our results,
thereby generating explicit expressions for the parameter $k$ and
the spectrum in terms of the dimension $d$.

On the basis of the freedom to apply conformal transformations
to the string action,
we conjecture that the issue of spacetime singularities
in general relativity may be solved by their stringy extensions.
An encounter with a spacetime singularity can be avoided
in a string theory by an appropriate choice of conformal gauge,
which does not change the local physics but which removes
the singularity.
We test this conjecture in the context of our cosmological
solutions in $d$ dimensions,
for which there are spacetime singularities associated
with the big bang/crunch.
It is indeed possible to avoid the usual problems
by a choice of conformal frame.

We expect generalizations of our solutions to exist.
For example,
a solution is likely to exist describing a background
with a non-zero condensate of the Kalb-Ramond field.
Similarly,
extensions of the equations to superstrings and
heterotic strings probably also permit consistent solutions
of the type discussed here.
In this context,
it would be particularly interesting to examine the
beta functions for the heterotic strings with tachyons,
in which tachyon condensates might be expected to form.

\vglue 0.6cm
{\bf \noindent 6. Acknowledgments}
\vglue 0.4cm

V.A.K. thanks Trinity College, Cambridge,
the Theory Division at CERN,
and the Aspen Center for Physics
for hospitality while part of this work was done.
This work was supported in part
by the North Atlantic Treaty Organization
under grant number CRG 910192
and by the United States Department
of Energy under contracts DE-AC02-84ER40125 and DE-FG02-91ER40661.

\vglue 0.6cm
{\bf\noindent 7. References}
\vglue 0.4cm


\begin{thebibliography}{xx}

\bibitem{CFMP}
C.G. Callan, D. Friedan, E.J. Martinec, and M.J. Perry,
Nucl. Phys. B 262 (1985) 593.

\bibitem{FT}
E.S. Fradkin and A.A. Tseytlin,
Nucl. Phys. B 261 (1985) 11.

\bibitem{L}
C. Lovelace,
Nucl. Phys. B 273 (1986) 413.

\bibitem{CKP}
C. Callan, I. Klebanov, and M. Perry,
Nucl. Phys. B 278 (1986) 78.

\bibitem{BNY}
R. Brustein, D. Nemeschansky, and S. Yankielowicz,
Nucl. Phys. B 301 (1988) 224.

\bibitem{CG}
C.G. Callan and Z. Gan,
Nucl. Phys. B 272 (1986) 647.

\bibitem{DS}
S.R. Das and B. Sathiapalan,
Phys. Rev. Lett. 56 (1986) 2664.

\bibitem{T}
A.A. Tseytlin,
Phys. Lett. B 264 (1991) 311.

\bibitem{MY}
R.C. Myers, Phys. Lett. B 199 (1987) 371.

\bibitem{ABEN}
I. Antoniadis, C. Bachas, J. Ellis and D. Nanopoulos,
Phys. Lett. B 211 (1988) 393;
Nucl. Phys. B328 (1989) 117;
Phys. Lett. B 257 (1991) 278.

\bibitem{APS}
S.P de Alwis, J. Polchinski, and R. Schimmrigk,
Phys. Lett. B 218 (1989) 449.

\bibitem{P}
J. Polchinski,
Nucl. Phys. B 346 (1990) 253.

\bibitem{M}
M. Mueller,
Nucl. Phys. B337 (1990) 37.

\bibitem{FF}
B.L. Feigin and D.B. Fuks,
Funct. Anal. Appl. 16 (1982) 114.

\bibitem{WZW}
J. Wess and B. Zumino,
Phys. Lett. B 37 (1971) 95;
E. Witten, Commun. Math. Phys. 92 (1984) 455.

\bibitem{BS}
I. Bars and K. Sfetsos,
preprint UCS-92/HEP-B1, hep-th/9205037;
I. Bars,
Phys. Lett. B 293 (1992) 315.

\bibitem{NW}
C. Nappi and E. Witten,
Phys. Lett. B 293 (1992) 309.

\bibitem{GQ}
P. Ginsparg and F. Quevedo,
Nucl. Phys. B385 (1992) 527.

\bibitem{KS2}
V.A. Kosteleck\'y and S. Samuel,
Phys. Rev. Lett. 63 (1990) 224;
Nucl. Phys. B336 (1990) 263.

\bibitem{KKS}
T. Kugo and K. Suehiro,
Nucl. Phys. B337 (1990) 434;
T. Kugo, H. Kunimoto, and K. Suehiro,
Phys. Lett. B 226 (1989) 48.

\bibitem{SZ}
M. Saadi and B. Zweibach,
Ann. Phys. 192 (1989) 213;
B. Zweibach, preprint IASSNS-HEP-92/41, hep-th/9206084.

\bibitem{K}
M. Kaku,
Phys. Rev. D 41 (1990) 3734.

\bibitem{KS1}
V.A. Kosteleck\'y and S. Samuel,
Phys. Rev. D 42 (1990) 1289.

\bibitem{CT}
C.B. Thorn,
Nucl. Phys. B 248 (1984) 551.

\bibitem{NY}
D. Nemeschansky and S. Yankielowicz,
Phys. Rev. Lett. 54 (1985) 620.

\bibitem{SI}
W. Siegel,
Phys. Lett. B 142 (1984) 276;
151 (1984) 391, 396.

\bibitem{KS3}
V.A. Kosteleck\'y and S. Samuel,
Phys. Rev. D 39 (1989) 683.

\bibitem{TM}
J. Thierry-Mieg,
Phys. Lett. B 171 (1986) 163.

\end{thebibliography}
\end{document}